\documentclass[aps, prl, reprint, superscriptaddress]{revtex4-2}

\usepackage{helvet}
\usepackage[utf8]{inputenc}
\usepackage{amsmath}
\usepackage{amssymb}
\usepackage{hyperref}
\usepackage{graphicx}
\usepackage{epstopdf}
\usepackage{dcolumn}
\usepackage{soul}
\usepackage{bm}
\usepackage{soul}
\usepackage{xspace}
\usepackage{verbatim}
\usepackage{color}
\usepackage{xcolor}

\hypersetup{colorlinks=true,linkcolor=blue,citecolor=blue, filecolor=blue,urlcolor=blue,breaklinks=true}

\begin{document}

\title{Quantum Metrology with Boundary Time Crystals}

\author{Victor Montenegro}
\email{vmontenegro@uestc.edu.cn}
\affiliation{Institute of Fundamental and Frontier Sciences, University of Electronic Science and Technology of China, Chengdu 610051, China}

\author{Marco G. Genoni}
\email{marco.genoni@fisica.unimi.it}
\affiliation{Quantum Technology Lab $\&$ Applied Quantum Mechanics Group, Dipartimento di Fisica ``Aldo Pontremoli'', Universit\`a degli Studi di Milano, I-20133 Milano, Italia}

\author{Abolfazl Bayat}
\email{abolfazl.bayat@uestc.edu.cn}
\affiliation{Institute of Fundamental and Frontier Sciences, University of Electronic Science and Technology of China, Chengdu 610051, China}

\author{Matteo G. A. Paris}
\email{matteo.paris@fisica.unimi.it}
\affiliation{Quantum Technology Lab $\&$ Applied Quantum Mechanics Group, Dipartimento di Fisica ``Aldo Pontremoli'', Universit\`a degli Studi di Milano, I-20133 Milano, Italia}

\date{\today}

\begin{abstract}
\noindent Quantum sensing is one of the arenas that exemplifies the superiority of quantum technologies over their classical counterparts. Such superiority, however, can be diminished due to unavoidable noise and decoherence of the probe. Thus, metrological strategies to fight against or profit from decoherence are highly desirable. This is the case of certain types of decoherence-driven many-body systems supporting dissipative phase transitions, which might be helpful for sensing. Boundary time crystals are exotic dissipative phases of matter in which the time-translational symmetry is broken, and long-lasting oscillations emerge in open quantum systems at the thermodynamic limit. We show that the transition from a symmetry unbroken into a boundary time crystal phase, described by a second-order transition, reveals quantum-enhanced sensitivity quantified through quantum Fisher information. We also determine the critical exponents of the system and establish their relationship. Our scheme is indeed a demonstration of harnessing decoherence for achieving quantum-enhanced sensitivity. From a practical perspective, it has the advantage of being independent of initialization and can be captured by a simple measurement.
\end{abstract}

\maketitle

\noindent \textbf{\large{Introduction}}

\noindent Quantum metrology protocols promise to achieve higher precision in the estimation of physical parameters compared with their classical counterparts~\cite{Giovannetti2004, Giovannetti2006, Giovannetti2011,Degen}, with applications ranging from biology~\cite{Taylor2016}, optical interferometry~\cite{Caves1981,RafalPO}, photonics and imaging~\cite{Pirandola:2018aa,Albarelli-2020}. 
One of the main issues in realizing quantum metrology protocols is the preparation of resourceful quantum probes. Besides strategies based on measurement and/or quantum control~\cite{WisemanMilburn,Geremia2003,Burgarth2015,Pang2017,Albarelli2017a,Sekatski2017quantummetrology,Albarelli2018restoringheisenberg,Rossi2020PRL,Zhou2018,Liu2017,Lin2021,Fallani2022,Tratzmiller2020,Montenegro2021,Montenegro2022seq}, a promising avenue is given by exploiting critical quantum systems. Two possibilities have been explored: (i) the ground state of critical Hamiltonians; and (ii) many-body systems with dissipative phase transitions. In the former, the ground state of critical Hamiltonians becomes highly sensitive with respect to the parameters driving the phase transition when approaching criticality~\cite{zanardi2007critical,zanardi2008quantum, invernizzi2008optimal,Tsang2013,salvatori2014quantum,Horodecki2018prx,Garbe2020,CAROLLO20201,Jianming2021critical,Liu2021,Mishra2021, Mishra2022,Ding2022,Sarkar2022topological,PhysRevA.105.042620}.
In the latter, dissipative phase transitions occur via a gap closing in the  spectrum of the Liouvillian describing the open system dynamics~\cite{sieberer2016keldysh, minganti2018spectral, lledo2019driven}. In this case, the steady-states present a divergent susceptibility with respect to one or more parameters characterizing the system evolution. This allows to exploit dissipative driven phase transitions for metrology purposes, in the presence of symmetry-breaking~\cite{Fernandez2017}, with Kerr resonators~\cite{Heugel2019,Dicandia2021}, with a finite-component system~\cite{Garbe2020,ivanov2020enhanced}, and via continuous measurements~\cite{Ilias2022, Cabot2022, cabot2023continuous}.

The breaking of spatial symmetry results in the existence of crystals. In a seminal paper, Wilczek~\cite{Wilczek2012} predicted that breaking temporal symmetry might also be possible, leading to the emergence of time crystals~\cite{Shapere2012, Wilczek2012, Tongcang2012}. In many-body systems, this is manifested through long-lasting periodic oscillations of an order parameter, with zero decay at the thermodynamic limit~\cite{Sacha_2018-time-crystal-review}. For not-too-long-range interactions, time crystals cannot emerge in any system with energy being the only conserved quantity, such as the ground or Gibbs thermal states~\cite{Watanabe2015, PhysRevLett.111.070402}. In contrary, long-range interactions~\cite{PhysRevLett.123.210602}, density-dependent gauge fields~\cite{PhysRevLett.123.250402, PhysRevLett.124.178901, PhysRevResearch.2.032038, syrwid2021can}, and extensive dynamical symmetries~\cite{PhysRevB.102.041117} can facilitate their emergence. So far, time crystals have been identified for both discrete and continuous temporal symmetry breakings. The former, which has been investigated theoretically~\cite{Sakurai2021, Else2016, Guo_2020, Russomanno2017, Pizzi2021, Surace2019, Gambetta2019, Riera_Campeny_2020, Estarellas2020} and demonstrated experimentally~\cite{taheri2022all, Zhang2017, Choi2017, Pal2018, Smits2018, Rovny2018, Mi2022}, can be observed in periodically driven systems in which an order parameter oscillates with a multiple frequency of the driving field~\cite{Sacha_2018-time-crystal-review}. The latter, is identified through dissipative open quantum many-body systems, which includes both  bulk~\cite{booker2020non, kongkhambut2022observation, Bu_a_2019} and boundary time crystals (BTCs)~\cite{Iemini2018, Carollo2022, Lledo_2020}. In contrast to dissipative phase transitions, where the Liouvillian gap closes for both the real and imaginary parts, for BTCs the real part closes while the imaginary part forms band gaps~\cite{minganti2018spectral, Iemini2018, Minganti2020}. This leads to a distinctive feature of BTCs: persistent oscillations in their stationary dynamics~\cite{Iemini2018, alaeian2022exact}.

The paradigmatic example of a BTC involves a simultaneous collective driving and dissipation of a system described by a large spin~\cite{Iemini2018}. See related works on BTCs in generalized systems~\cite{PhysRevB.103.184308, piccitto2021symmetries}. This model was extensively studied for its quantum optical properties~\cite{Agarwal1977,Carmichael1977}, along with the critical behavior of the steady-state of such dynamics~\cite{Walls1978,PURI1979200,Walls1980,Carmichael1980,Morrison2008}, whose signatures have been recently observed experimentally for a small effective atom number~\cite{Ferioli2023}. Mean-field analysis of this BTC phase transition has been also put forward~\cite{Carollo2022}, the possibility of distinguishing the two phases via continuous monitoring~\cite{Cabot2022, cabot2023continuous}, and the study of the many-body quantum correlations building up in the two phases~\cite{Lourenco2022}. Several open problems still exist: (i) despite analyses that identify the BTC transition as a second-order type, its  critical features (e.g., critical exponents) have hardly been investigated; (ii) the possibility of BTC transition as a resource for quantum sensing has not yet been explored; and (iii) whether a simple physical measurement can reveal the BTC enhanced sensitivity.

In this work, through several finite-size scaling analyses, we show that the transition from symmetry unbroken into a boundary time crystal phase can indeed be exploited for quantum-enhanced sensitivity. This is evidenced as the corresponding quantum Fisher information (QFI) presents a super-classical scaling $N^b$ with $b>1$, in terms of the probe size $N$, and thus overcoming the so-called standard quantum limit~\cite{Giovannetti2004,Giovannetti2006,Giovannetti2011}. We find the corresponding critical exponents for the QFI through independent finite-size scaling analyses and established an equality among them confirmed by our numerical simulations.  This provides further confirmation for the validity of our analysis. Finally, we show that a simple measurement can achieve the aforementioned enhanced precision.

\hfill

\noindent \textbf{\large{Results}}

\noindent \textbf{Quantum parameter estimation}

\noindent In this section, we briefly review the parameter estimation theory. Quantum parameter estimation aims to infer an unknown quantity $\omega$ encoded in the quantum state of a probe $\rho_\omega$ by performing a proper measurement~\cite{Matteo2009}. For a given measurement described by a set of positive operator-valued measure (POVM) $\{\Pi_s\}$, each outcome $s$ appears with the probability $p(s|\omega){=}\mathrm{Tr}[\Pi_s\rho_\omega]$. The Cram\'{e}r-Rao inequality sets a fundamental bound for the estimation of $\omega$ for the given POVM as $\mathrm{Var}[\omega]{\geq}\mathcal{F}_C(\omega)^{-1}$~\cite{Cramer-1946, Braunstein-1994}, where $\mathrm{Var}[\omega]$ is the variance of the estimation and $\mathcal{F}_C(\omega){=}\sum_{s}p(s|\omega)^{-1} [\partial_\omega p(s|\omega)]^2$ ($\partial{/}\partial\omega{:=}\partial_\omega$) is the classical Fisher information (CFI). 
Bayesian estimator and Maximum Likelihood estimator are proven to be optimal, i.e. to saturate the Cram\'er-Rao bound, in the asymptotic limit of large number of measurements, while their near-optimality properties have been extensively shown in several experimental instances also in the more practical regime of finite number of measurements~\cite{BrivioPRA, GenoniPhDiffCoh, BlandinoPRL}. One can then optimize over all possible POVMs to achieve the ultimate precision limit determined by quantum Fisher information (QFI) $\mathcal{F}_Q(\omega){=}\max_{ \{\Pi_s\}}\mathcal{F}_C(\omega)$, resulting in a tighter bound of the Cram\'{e}r-Rao inequality $\mathrm{Var}[\omega]{\geq}\mathcal{F}_C(\omega)^{-1}{\geq}\mathcal{F}_Q(\omega)^{-1}$. While several expressions exist for computing the QFI, we use throughout our work
\begin{equation}
    \mathcal{F}_Q(\omega) = 8 \lim_{\delta \omega\rightarrow 0} \frac{1 - \mathbb{F}(\rho_{\omega - \delta \omega},\rho_{\omega + \delta \omega})}{(2\delta \omega)^2},\label{eq:QFI_main_text}
\end{equation}
where
\begin{equation}
    \mathbb{F}(\rho_1, \rho_2) = \mathrm{Tr}\left[ \sqrt{ \sqrt{\rho_1} \rho_2 \sqrt{\rho_1} } \right],
\end{equation}
is the Fidelity between quantum states $\rho_1$ and $\rho_2$. Note that the QFI represents the ultimate theoretical sensing precision for a given probe. This benchmark is of utmost importance as it determines both the performance achieved by a specific measurement setup and establishes quantum-enhanced sensitivity concerning a given sensing resource. Both cases are elaborated in detail later in our work.

\hfill

\noindent \textbf{The model}

\noindent We consider a system of $N$ non-interacting spin-1/2 particles forming a pseudospin of length $S{=}N/2$. The collective angular momentum operators are given by $\hat{S}_\alpha{=}1{/}2\sum_j\sigma_\alpha^{(j)}$, where $\sigma_\alpha^{(j)}$ ($\alpha{=}x,y,z$) is the Pauli matrix at site $j$. Conventionally, one can define $\hat{S}_\pm{=}\hat{S}_x{\pm}iS_y$, satisfying $[\hat{S}_+,\hat{S}_-]{=}2\hat{S}_z$, $[\hat{S}_z,\hat{S}_\pm]{=}\hat{S}_\pm$. We consider the Hamiltonian of the system to be $H{=}\omega \hat{S}_x$, where $\omega$ is the single particle coherent splitting. The evolution of the open system with collective spin dissipation is given by the Lindbladian master equation
\begin{equation}
\frac{d}{dt}\rho = -i\omega[\hat{S}_x,\rho]+\frac{\kappa}{S}\left(\hat{S}_-\rho \hat{S}_+-\frac{1}{2}\left\{\hat{S}_+\hat{S}_-,\rho\right\} \right)=\mathcal{L}[\rho],\label{eq:master-equation}
\end{equation}
where $\mathcal{L}[\rho]$ is the Liouvillian and $\kappa$ is the collective dissipation rate. Eq.~\eqref{eq:master-equation} has also been studied in the presence of local pumping and local anisotropies in the coherent splitting parameter, accurately describing some experimental setups~\cite{Norcia_2018, Shankar_2017}. In such scenarios, the temporal symmetry breaking still survives for a wide range of noise strengths~\cite{Tucker_2018}. One can interpret the origin of this master equation by the interaction between our system of $N$ particles sitting at the boundary of a large bulk with $N'$ particles~\cite{Iemini2018}. This implies that in the thermodynamic limit where both $N',N{\rightarrow}\infty$, the ratio $N/N'{\rightarrow}0$. The evolution of the boundary and the bulk is governed by a unitary operation. By tracing out the bulk degrees of freedom, one gets Eq.~\eqref{eq:master-equation}. At any time $t$, the density matrix of the boundary is given by $\rho(t){=}e^{\mathcal{L}t}\rho(0)$. As $\omega{/}\kappa$ varies, the steady state $\rho_\mathrm{SS}{=}\rho(t{\rightarrow}\infty)$ of the boundary goes through a phase transition from a static phase (determined by $\omega{<}\kappa$) into a BTC phase with long-lasting total spin oscillations (determined by $\omega{>}\kappa$). In the thermodynamic limit, the transition is characterized by a spontaneous temporal symmetry breaking at the transition point $\omega_c{=}\kappa$~\cite{Iemini2018}. By numerically solving~\eqref{eq:master-equation}~\cite{JOHANSSON20131234}, we focus on sensing the value of $\omega{/}\kappa$ from the steady state $\rho_\mathrm{SS}$ across the whole phase diagram.

\begin{figure}
%\centering \includegraphics[width=\linewidth]{fig1:expectation_S_v2.pdf}
%\centering \includegraphics[width=\linewidth]{fig1:spectra_liouvillian_v2.pdf}
\centering \includegraphics[width=\linewidth]{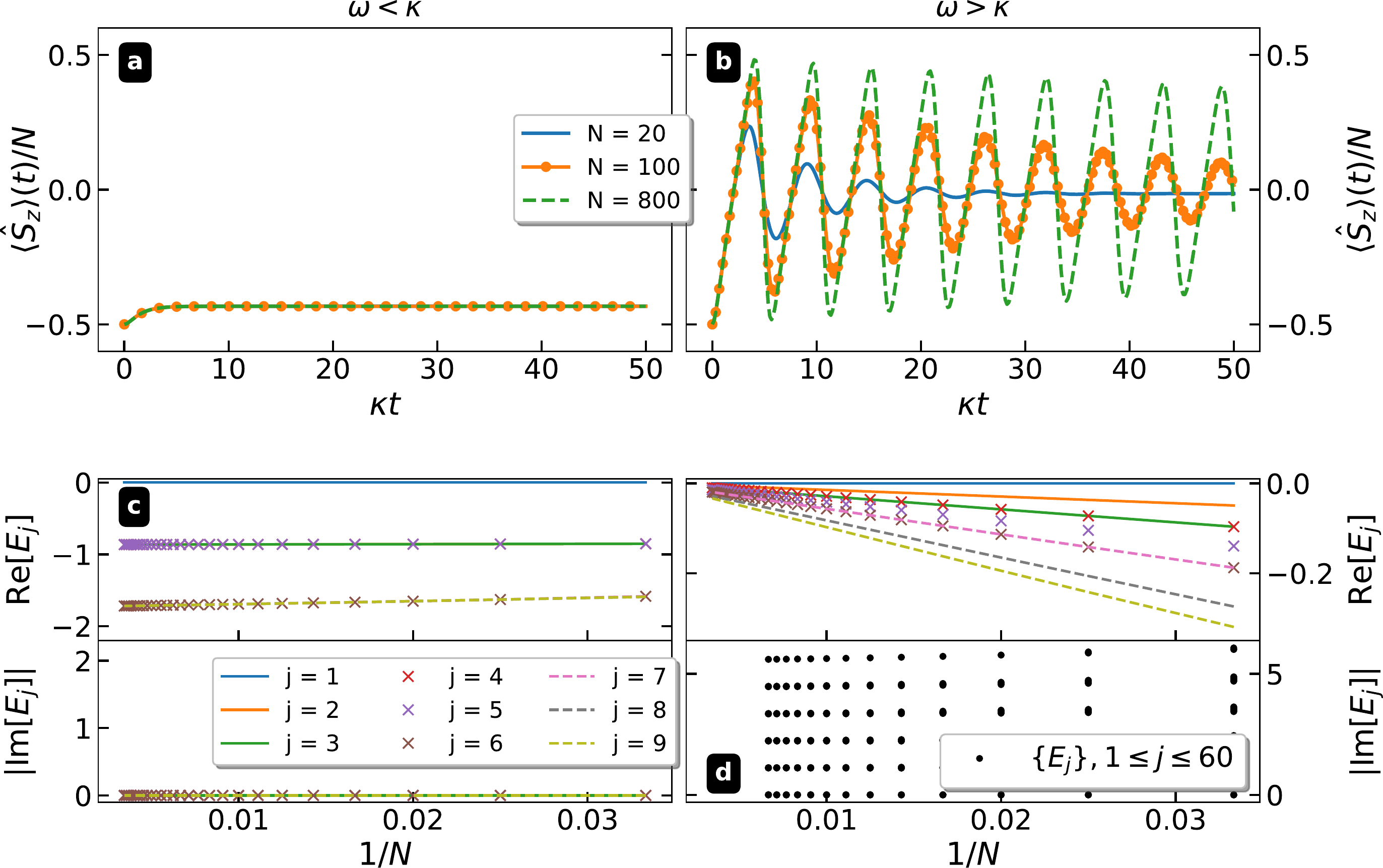}
\caption{\textbf{Boundary time crystal phases.} Comparison between the static phase (left column, $\omega{=}0.5\kappa$) and the BTC phase (right column, $\omega{=}1.5\kappa$). Panels \textbf{(a)} and \textbf{(b)} show the $z-$component of the total spin $\langle\hat{S}_z\rangle(t){/}N$ as a function of time $\kappa t$ for several systems sizes $N$. Panels \textbf{(c)} and \textbf{(d)} show the real and imaginary parts of Liouvillian's eigenvalues $E_j$ as a function of $1/N$.}\label{fig1:spectra_liouvillian}
\end{figure}

\hfill

\noindent \textbf{Boundary time crystals}

\noindent This section summarizes key features of the BTCs~\cite{Iemini2018}. To show the dynamics of the system in two phases, in Fig.~\ref{fig1:spectra_liouvillian}(a), we depict the $z-$component of the total spin $\langle\hat{S}_z\rangle(t){/}N$ as a function of time $\kappa t$ for several systems sizes $N$ in the symmetry unbroken phase ($\omega{<}\kappa$). The evolution is size independent, reaching its steady state without showing any oscillation. In contrast, as shown in Fig.~\ref{fig1:spectra_liouvillian}(b), in the BTC phase ($\omega{>}\kappa$), the system shows persistence oscillations and decay gets weaker as the system size increases. This suggests that in the thermodynamic limit, the oscillations perdure indefinitely. To understand the behavior of the static and the BTC phases, one has to investigate the Liouvillian eigenvalues. In Fig.~\ref{fig1:spectra_liouvillian}(c), we plot the nine most relevant eigenvalues of the Liouvillian, i.e., those with the lowest decaying rate due to smaller real values, in the static phase for various system sizes. These eigenvalues are real and non-positive, with one of them being zero determining a unique static steady state. The eigenvalues with imaginary parts (not shown in the figure) appear only with large negative real values and, thus, decay very fast. In contrast, as shown in Fig.~\ref{fig1:spectra_liouvillian}(d), the imaginary part of the eigenvalues form almost equally separated bands in the BTC phase, while in the thermodynamic limit, the real part of the eigenvalues goes to zero. The vanishing real part of the eigenvalues describes the slowing down of the decay as the system size increases. On the other hand, the frequency of the persistent oscillation is determined by the value of the almost equally separated bands of the imaginary part of Liouvillian's eigenvalues.

\hfill

\noindent \textbf{Characterization of the transition}

\begin{figure}
%\centering \includegraphics[width=\linewidth]{fig2:FSS_Magnetization_v2.pdf}
\centering \includegraphics[width=\linewidth]{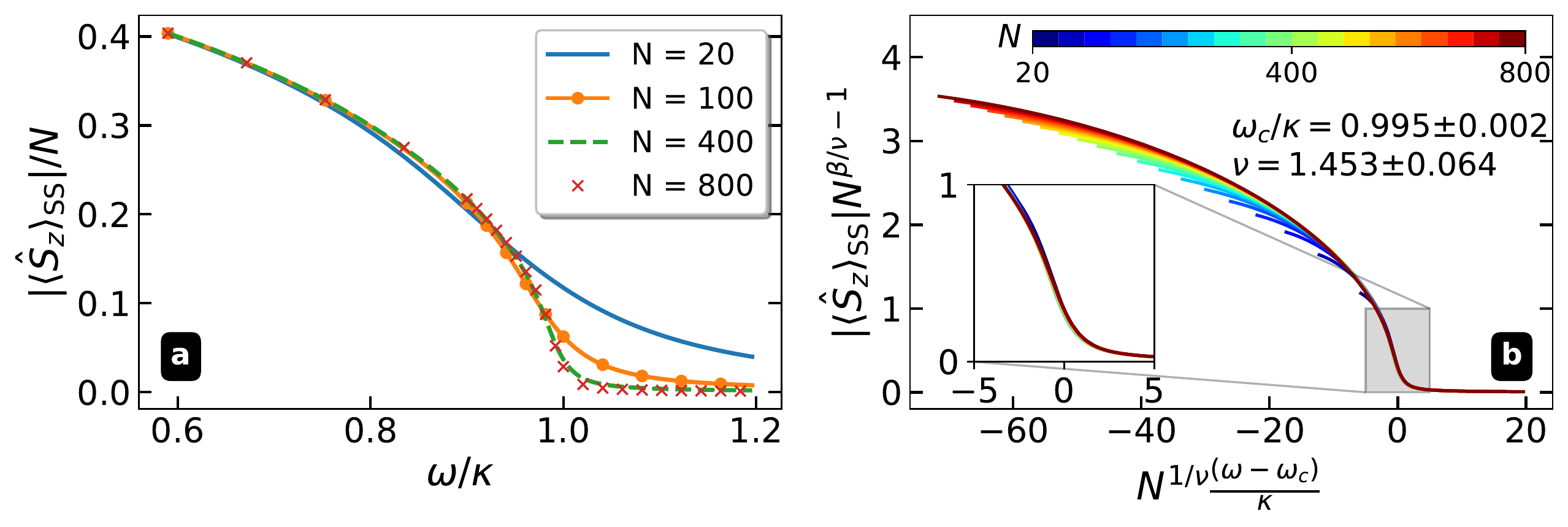}
\caption{\textbf{Magnetization finite-size scaling analysis. (a)} Average steady-state magnetization $|\langle\hat{S}_z\rangle_\mathrm{SS}|{/}N$ as a function of $\omega{/}\kappa$ for several system sizes $N$. \textbf{(b)} Finite-size scaling analysis, we plot $|\langle\hat{S}_z\rangle_\mathrm{SS}|N^{\frac{\beta}{\nu} - 1}$ as a function of $N^{\frac{1}{\nu}}(\omega{-}\omega_c){/}\kappa$ for various system sizes from $N{=}20$ to $N{=}800$.}\label{fig2:FSS_magnetization}
\end{figure}

\noindent To characterize the phase transition that occurs for the steady state, one can investigate the average steady-state magnetization, namely $\langle\hat{S}_z\rangle_\mathrm{SS}{=}\langle\hat{S}_z\rangle(t{\rightarrow}\infty)$. In the static phase ($\omega{<}\kappa$), the steady state magnetization $\langle\hat{S}_z\rangle_\mathrm{SS}$ takes non-zero values. In the BTC phase ($\omega{>}\kappa$), however, the $\langle\hat{S}_z\rangle(t)$ shows decaying oscillations around its steady state value $\langle\hat{S}_z\rangle_\mathrm{SS}$, which tends to zero in thermodynamic limit, i.e. $\lim_{N\rightarrow\infty} \langle\hat{S}_z\rangle_\mathrm{SS}{=}0 $. In the thermodynamic limit, the decay of $\langle\hat{S}_z\rangle(t)$ is suppressed, and long-lived oscillations persist with their time average being zero. In Fig.~\ref{fig2:FSS_magnetization}(a), we plot $|\langle\hat{S}_z\rangle_\mathrm{SS}|{/}N$ as a function of $\omega{/}\kappa$ for several system sizes $N$. As the system size increases, the transition from non-zero $\langle\hat{S}_z\rangle_\mathrm{SS}$ in the static phase into zero value in the BTC phase becomes sharper, suggesting a non-analytic behavior at the thermodynamic limit. This strongly hints that the transition might be second-order with $\langle\hat{S}_z\rangle_\mathrm{SS}$ playing the role of an order parameter. In the thermodynamic limit, a second-order phase transition near the critical point is expected to be described by an algebraically vanishing order parameter, i.e. $\langle\hat{S}_z\rangle_\mathrm{SS}{\sim}\left|\frac{\omega{-}\omega_c}{\kappa}\right|^{\beta}$ which is accompanied by the emergence of a diverging length scale $\xi{\sim}\left|\frac{\omega{-}\omega_c}{\kappa}\right|^{-\nu}$. The parameters $\nu$ and $\beta$ are critical exponents defining the transition's universality class. For finite-size systems, the order parameter gets some corrections and thus follows a conventional ansatz~\cite{newmanb99, PhysRevLett.28.1516}
\begin{equation}
\frac{|\langle\hat{S}_z\rangle_\mathrm{SS}|}{N}{=}N^{-\frac{\beta}{\nu}}f\left(N^{\frac{1}{\nu}}\frac{(\omega{-}\omega_c)}{\kappa}\right).
\end{equation}
To determine the critical exponents $\beta$ and $\nu$, in Fig.~\ref{fig2:FSS_magnetization}(b), we plot $|\langle\hat{S}_z\rangle_\mathrm{SS}|N^{\frac{\beta}{\nu} - 1}$ as a function of $N^{1/\nu}(\omega{-}\omega_c)/\kappa$ for various system sizes from $N{=}20$ to $N{=}800$. With the help of the Python package pyfssa~\cite{andreas_sorge_2015_35293, https://doi.org/10.48550/arxiv.0910.5403}, we tune the critical point $\omega_c$ and the exponents $\nu$ and $\beta$ such that curves from different system sizes collapse on each other around the critical point. Our analysis shows that $\omega_c{/}\kappa{=}0.995{\pm}0.002$, $\nu{=}1.453{\pm}0.064$, and $\beta{=}0.434{\pm}0.055$. The collapse of curves with different system sizes shows that the transition is of the second-order, and $\langle\hat{S}_z\rangle_\mathrm{SS}$ can characterize the transition. 

\hfill

\noindent \textbf{Boundary time crystal sensor}

\begin{figure}
% \centering \includegraphics[width=\linewidth]{fig3:QFI_scaling_top_v2.pdf}
% \centering \includegraphics[width=\linewidth]{fig3:QFI_scaling_bottom_v2.pdf}
\centering \includegraphics[width=\linewidth]{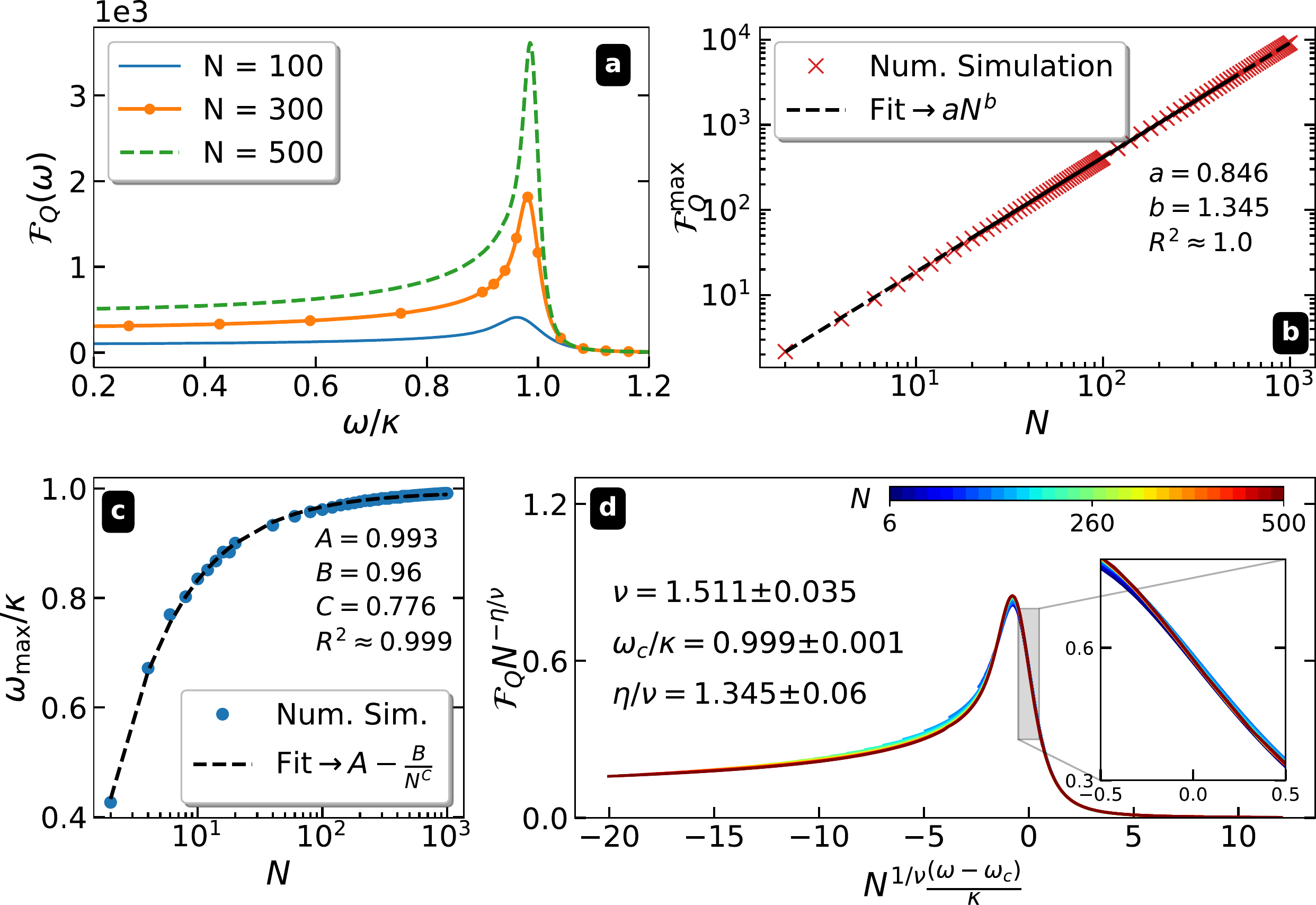}
\caption{\textbf{Quantum-enhanced sensitivity. (a)} Quantum Fisher information (QFI) $\mathcal{F}_Q(\omega)$ as a function of $\omega{/}\kappa$ for various system sizes $N$. \textbf{(b)} Peak of the QFI $\mathcal{F}_Q^\mathrm{max}{=}\mathcal{F}_Q(\omega_\mathrm{max})$ as a function of $N$. We fit a function of the form $\mathcal{F}_Q^\mathrm{max}\approx aN^{b}$, with $a{=}0.846$ and $b{=}1.345$. The coefficient $b{>}1$ evidences quantum-enhanced sensitivity. \textbf{(c)} $\omega_\mathrm{max}{/}\kappa$ as a function of $N$, we fit a Pareto function of the form $\omega_\mathrm{max}{=}\kappa(1{-}\frac{1}{N^{0.776}})$. \textbf{(d)} We plot $\mathcal{F}_Q N^{-\eta/\nu}$ as a function of $N^{1/\nu}(\omega{-}\omega_c)/\kappa$ for various system sizes from $N=6$ to $N=500$.}\label{fig3:scaling_QFI}
\end{figure}

\noindent While the presence of decoherence is mostly destructive on the sensing power of quantum probes~\cite{Escher2011,Demkowicz2012}, some specific types of decoherence which lead to dissipative phase transitions might be useful for sensing~\cite{Fernandez2017,Heugel2019,Dicandia2021,Garbe2020,ivanov2020enhanced,Ilias2022}.
The dissipative BTC phases show a second-order transition behavior~\cite{Agarwal1977,Carmichael1977,Walls1978,Walls1980,Carmichael1980,Morrison2008} which makes them even more interesting from a quantum metrology perspective. To investigate the sensing capacity of our BTC probe for estimating $\omega{/}\kappa$, in Fig~\ref{fig3:scaling_QFI}(a), we plot the QFI $\mathcal{F}_Q(\omega)$ as a function of $\omega{/}\kappa$ for various system sizes. Fig~\ref{fig3:scaling_QFI}(a) resembles the one obtained in~\cite{Lourenco2022}, where the QFI was evaluated for a generic spin rotation to characterize the quantum correlations in the two phases. Two interesting features can be observed: (i) QFI indeed shows a peak near the transition point; and (ii) the point at which the QFI peaks, i.e., $\omega{=}\omega_\mathrm{max}$, shifts toward $\omega_c{=}\kappa$ as the system size increases. By taking the peak of the QFI $\mathcal{F}_Q^\mathrm{max}{=}\mathcal{F}_Q(\omega_\mathrm{max})$, one can investigate the scaling with respect to the probe size $N$ in order to identify a possible quantum-enhanced precision. In Fig.~\ref{fig3:scaling_QFI}(b), we plot $\mathcal{F}_Q^\mathrm{max}$ as a function of $N$, which can be precisely mapped by a fitting function $\mathcal{F}_Q^\mathrm{max}{\approx}aN^{b}$, with $a{=}0.846$ and $b{=}1.345$. Clearly, $\mathcal{F}_Q^\mathrm{max}$ shows quantum-enhanced sensitivity, i.e. super-linear scaling surpassing the standard quantum limit, as it diverges with the exponent $b{>}1$ by increasing the system size. Note that the decoherence induces the BTC phase transition and thus contributes to achieving quantum-enhanced sensitivity. In Fig.~\ref{fig3:scaling_QFI}(c), we plot $\omega_\mathrm{max}{/}\kappa$ as a function of $N$, which shows asymptotic convergence towards $\omega_c{=}\kappa$ through a fitting function of the form $\omega_\mathrm{max}{=}\kappa(1{-}N^{-0.776})$.

These analyses show that $\mathcal{F}_Q$ should follow an ansatz of the following type
\begin{equation}
\mathcal{F}_Q(\omega) = \frac{a}{N^{-b}+c\left(\frac{\omega - \omega_\mathrm{max}(N)}{\kappa}\right)^\eta},\label{eq:ansazt1}
\end{equation}
for some constants $a,b,c$ and $\eta$. At $\omega{=}\omega_\mathrm{max}$, one can retrieve $\mathcal{F}_Q^\mathrm{max}{\sim}aN^b$. On the other hand, for $N{\rightarrow}\infty$, one can get $\mathcal{F}_Q(\omega){\sim}\left|\frac{\omega{-}\omega_c}{\kappa}\right|^{-\eta}$. To estimate $\eta$, one has to perform a finite-size scaling analysis.
\begin{equation}
\mathcal{F}_Q(\omega)=N^{\frac{\eta}{\nu}}f\left(N^{\frac{1}{\nu}}\frac{(\omega-\omega_c)}{\kappa}\right).\label{eq:ansazt2}
\end{equation}
The second-order nature of the transition implies that all quantities, including QFI, should show scale invariance near the transition point. In Fig.~\ref{fig3:scaling_QFI}(d), we plot $\mathcal{F}_Q N^{-\eta/\nu}$ as a function of $N^{1/\nu}(\omega{-}\omega_c)/\kappa$ for various system sizes from $N{=}6$ to $N{=}500$. Using pyfssa~\cite{andreas_sorge_2015_35293, https://doi.org/10.48550/arxiv.0910.5403}, we determine the critical point $\omega_c{/}\kappa{=}0.999{\pm}0.001$, the critical exponent $\eta{=}2.031{\pm}0.043$, and $\nu{=}1.511{\pm}0.035$. First, $\omega_c$ and $\nu$ determined from the finite-size scaling analysis of the plot of the QFI are very close to the ones from $\langle\hat{S}_z\rangle_\mathrm{SS}$, showing the consistency of our analysis. Second, since both Eqs.~\eqref{eq:ansazt1} and~\eqref{eq:ansazt2} describe the QFI, they should be similar. In the limit of large $N$, where $\omega_\mathrm{max}(N){\simeq}\omega_c$ in Eq.~\eqref{eq:ansazt1}, a calculation shows that the two ansatzes are of the same form if $b{=}\eta/\nu$ (see Methods Section). Thus, the three critical exponents $b, \eta$, and $\nu$ are not independent. In fact, the values found for $\eta$ and $\nu$ from the finite-size scaling of Fig.~\ref{fig3:scaling_QFI}(d), perfectly matches with the exponent $b$ computed from an independent scaling analysis in Fig.~\ref{fig3:scaling_QFI}(b), i.e. $\eta/\nu{=}1.345 \pm 0.06$ where $b{=}1.345$.

\hfill

\noindent \textbf{Classical Fisher information}

\begin{figure}
% \centering \includegraphics[width=\linewidth]{fig4:CFI_top_v3.pdf}
% \centering \includegraphics[width=\linewidth]{fig4:CFI_bottom_v3.pdf}
\centering \includegraphics[width=\linewidth]{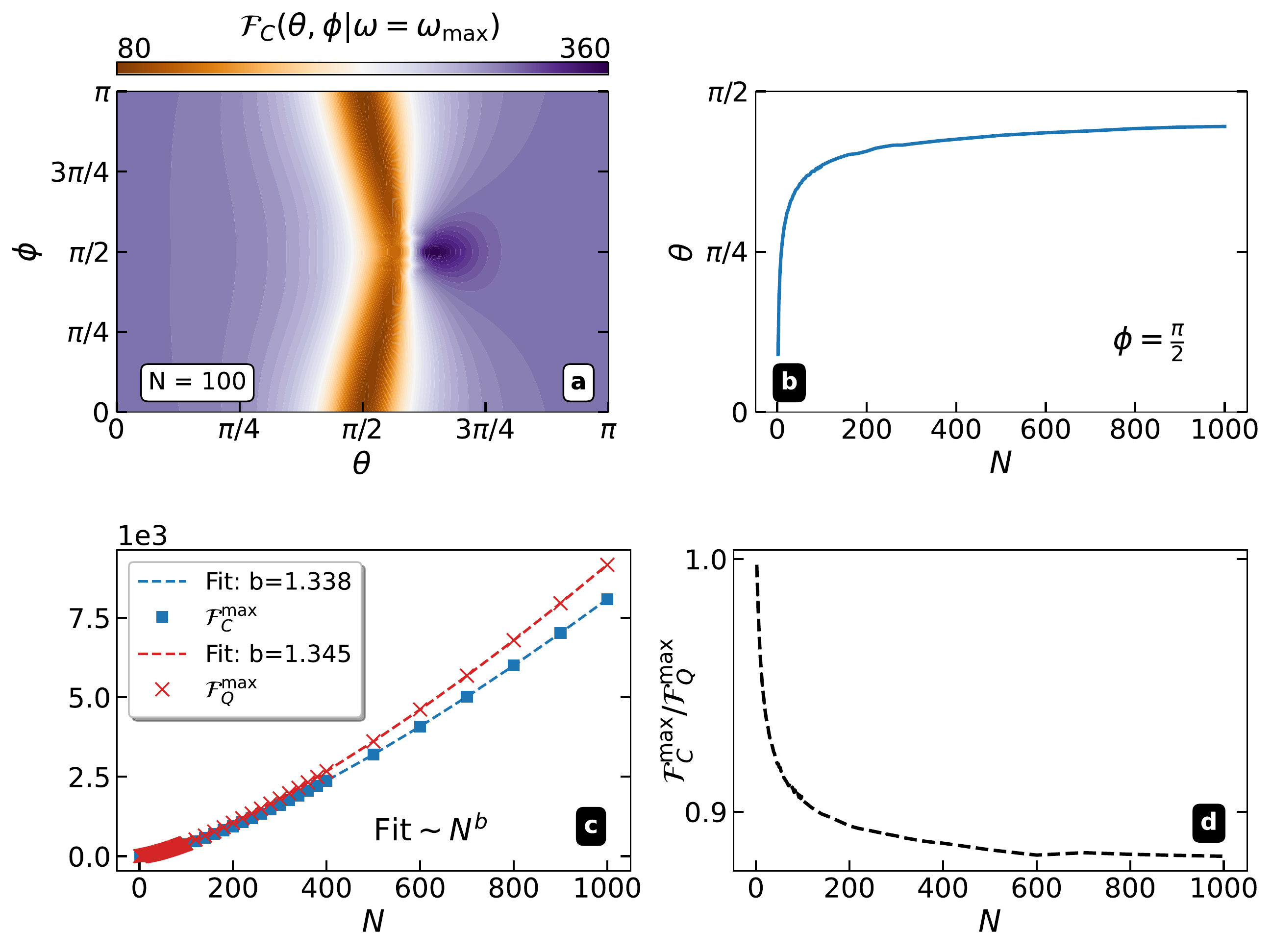}
\caption{\textbf{Sensing precision with a feasible measurement. (a)} Classical Fisher information (CFI) $\mathcal{F}_C(\theta,\phi|\omega{=}\omega_\mathrm{max})$ as functions of the angles $\theta$ and $\phi$ at $\omega{=}\omega_\mathrm{max}$ for $N=100$. \textbf{(b)} Dependence of the angles $\theta$ and $\phi$ as a function of the system size $N$. \textbf{(c)} We plot the CFI $\mathcal{F}_C^\mathrm{max}$ and the quantum Fisher information (QFI) $\mathcal{F}_Q^\mathrm{max}$ as a function of the system size $N$. We fit a function of the form ${\sim}N^b$, for both the CFI and the QFI, which evidences quantum-enhanced sensing (i.e., $b{>}1$) even for a suboptimal measurement. \textbf{(d)} Efficiency ratio $\mathcal{F}_C^\mathrm{max}{/}\mathcal{F}^\mathrm{max}_Q$ as a function of probe size $N$.}\label{fig5:CFI}
\end{figure}

\noindent As mentioned before, the optimal measurement basis that saturates the Cram\'{e}r-Rao inequality, in general, depends on the unknown parameter and is highly entangled, which makes it practically unfeasible. Hence, determining the estimation performance with a suboptimal yet available set of measurements is highly desirable. We consider the spin projection $\hat{S}_{\hat{\bm{n}}}{=}\hat{\bm{n}}{\cdot}\hat{\bm{S}}$, where $\hat{\bm{n}}{=}(\sin\theta\cos\phi{,}\sin\theta\sin\phi{,}\cos\theta)$ is the unit vector in spherical coordinates and $\hat{\bm{S}}{=}(\hat{S}_x{,}\hat{S}_y{,}\hat{S}_z)$. The eigenvectors of $\hat{S}_{\hat{\bm{n}}}$ are given by $|s\rangle{:=}|s(\theta,\phi)\rangle$ for given angles $\theta$ and $\phi$, such that $\hat{S}_{\hat{\bm{n}}}|s\rangle{=}s|s\rangle$, with $s$ taking values from $-N/2$ to $+N/2$. By measuring $\hat{S}_{\hat{\bm{n}}}$, every outcome appears with probability $p(s|\omega){=}\langle s|\rho_\mathrm{SS}|s\rangle$, and thus, one can get the corresponding CFI $\mathcal{F}_C(\omega)$. In Fig.~\ref{fig5:CFI}(a), we show the CFI $\mathcal{F}_C$ at $\omega{=}\omega_\mathrm{max}$ as functions of the rotated angles $\theta$ and $\phi$ for $N=100$. As the figure shows, a clear maximum of the CFI $\mathcal{F}_C^\mathrm{max}$ is obtained for some optimized $\theta$ and $\phi$ values. No extra information of the CFI for values $\pi{\leq}\phi{\leq}2\pi$ is found. In Fig.~\ref{fig5:CFI}(b), we plot the dependence of such optimized tuple $\theta$ and $\phi$ as a function of the system size $N$. As seen from the figure, a clear trend towards $\theta\rightarrow \pi/2$ for $N\gg 1$ is observed, whereas $\phi=\pi/2$. In Fig.~\ref{fig5:CFI}(c), we plot the CFI $\mathcal{F}_C^\mathrm{max}$ at $\omega{=}\omega_\mathrm{max}(N)$, optimized over the angles $\theta$ and $\phi$, as a function of the system size $N$. As seen from the figure, the CFI follows the QFI curve. Indeed, a fitting function $\mathcal{F}_C^\mathrm{max}{\sim}N^{1.338}$ reveals quantum-enhanced sensitivity concerning the system size $N$ even for the present suboptimal choice of the measurement. To quantify the efficiency of our simple measurement, in Fig.~\ref{fig5:CFI}(d), we plot the ratio $\mathcal{F}^\mathrm{max}/\mathcal{F}^\mathrm{max}_Q$ as a function of the system size $N$. As the figure shows, a fair fraction of the ultimate sensing performance can be extracted by the simple measurement $\hat{S}_{\hat{\bm{n}}}$.

\begin{figure}[t]
% \centering \includegraphics[width=\linewidth]{time-constrained-qfi.pdf}
\centering \includegraphics[width=\linewidth]{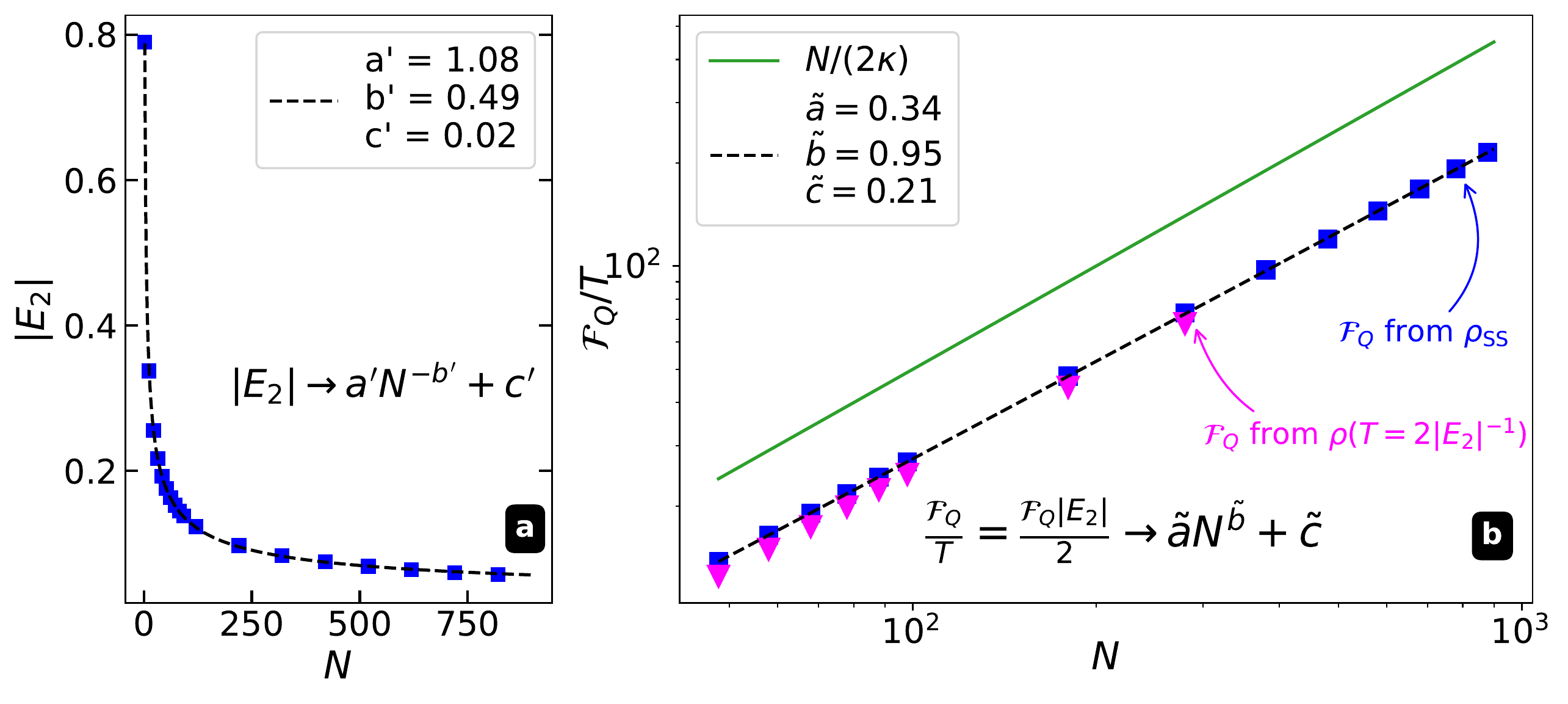}
\caption{\textbf{Time-constrained sensing bound. (a)} Absolute value of the second Liouvillian eigenvalue $|E_2|$ as a function of the system size $N$ at $\omega_\mathrm{max}$. We fit a function of the form $|E_2|{\sim}a'N^{-b'}{+}c'$ showing that $\tau^{-1}{\sim}|E_2|{\sim}N^{-0.49}$. \textbf{(b)} We plot different figures of merit as a function of the system size $N$. Magenta triangles: ratio $\mathcal{F}_Q(\omega)/T$ where $\mathcal{F}_Q(\omega)$ is the quantum Fisher information (QFI) computed from the quantum state evolved from an initial spin ground state up to a time $T{=}2\tau{=}2/|E_2|$; blue squares: ratio $\mathcal{F}_Q(\omega)/T$ where $\mathcal{F}_Q(\omega)$ is the QFI computed from the steady-state obtained at $\omega_\mathrm{max}$, divided by $T{=}2\tau$; green line: the upper bound $N/(2\kappa)$ reported in Eq.~\eqref{eq:DDbound}.}\label{SM_time_v2}
\end{figure}

\hfill

\noindent \textbf{Time-constrained sensing protocol}

\noindent 
We have shown how the BTC phase transition can be harnessed to obtain a quantum-enhanced super-linear scaling in the probe size $N$. This result has been obtained considering $N$ as the only resource, without putting any constraint on the time needed for running the metrology protocol; while this is a sensible assumption both from a fundamental and practical point of view, it is similarly important to consider scenarios where
the total time $T$ needed for accomplishing the sensing task has to
be incorporated into the resource analysis. In these cases, instead of $\mathcal{F}_Q$ one has to assess $\mathcal{F}_Q{/}T$ as the main figure of merit~\cite{Chin2012,Chaves2013,Brask2015,Smirne2016,Albarelli2018restoringheisenberg,Rossi2020PRL}. In criticality-based sensing scenarios, the relevant time $T$, that corresponds to the probes' state preparation, typically increases with $N$, and thus the scaling of $\mathcal{F}_Q{/}T$ may diminish in contrast to $\mathcal{F}_Q$~\cite{Horodecki2018prx,PhysRevA.105.042620,Garbe2020}. In our case, an upper bound for the QFI at fixed evolution time $T$ and probe size $N$ can be derived analytically~\cite{Rafal2017} (see Methods Section). We indeed obtain that the following inequality has to be always satisfied $\mathcal{F}_Q{/}T{\lesssim}N/2$, showing that eventually $\mathcal{F}_Q{/}T$ has to follow a linear scaling in $N$. However, note that the linear scaling of the ratio $\mathcal{F}_Q{/}T$ is a result of the quantum-enhanced sensitivity and should not be mistaken by standard limit of the QFI with respect to $N$. Indeed, for classical sensors in which the Fisher information scales linearly with $N$, the normalized $\mathcal{F}_Q{/}T$ will scale sub-linearly with the system size. In our protocol, the time $T$ needed to reach the steady-state is determined by the smallest real eigenvalue of the Liouvillian $\mathcal{L}$, which turns out to be the second eigenvalue $E_2$. In other words, the dominant decay of the observables is given by $\exp\{-|E_2|t\}$ (note that $E_2$ is real and negative), and we can thus identify a typical time scale $\tau = 1/|E_2|$ that rules the dynamics.

In Fig.~\ref{SM_time_v2}(a) we plot $|E_2|$ as a function of the system size $N$. As it is common in most of the critical metrology protocols, we find that the time needed to prepare the critical quantum state diverges with the probe size. In particular, we have $\tau \approx N^{b'}$ with a certain exponent $b'$ that we have obtained from a numerical fit. To properly assess the behavior of the figure of merit $\mathcal{F}_Q(\omega)/T$, in Fig.~\ref{SM_time_v2}(b), we have numerically simulated the dynamics up to a time $T{=}2\tau{=}2/|E_2|$ for different values of $N$ and evaluated the QFI of the corresponding quantum state. We first observe how these values are approximately equal to the ones obtained by evaluating the QFI from the critical steady-state divided by $T=2\tau$, confirming that at this evolution time, the steady-state is approximately obtained. Most importantly we observe that also for our protocol $\mathcal{F}_Q(\omega)/T$ follows an approximate linear scaling in the probe size $N$, with a reduced constant factor respect to the (typically not achievable) upper bound (\ref{eq:DDbound}).

\hfill

\noindent \textbf{\large{Discussion}}

\noindent Decoherence is highly detrimental for a large class of quantum-enhanced sensing protocols~\cite{Escher2011,Demkowicz2012}. In practice, such enhancement is already very challenging to achieve, and a non-classical scaling of the precision may be recovered only for particular instances of noise or via certain quantum control strategies~\cite{Chin2012,Chaves2013,Brask2015,Smirne2016,Sekatski2017quantummetrology,Zhou2018,Albarelli2018restoringheisenberg,Rossi2020PRL}. In this work, we have proposed a boundary time crystal phase transition for harnessing decoherence to achieve quantum-enhanced sensitivity. Our procedure demonstrates a decoherence-induced sensing scheme and sheds light on the exotic boundary time crystal phase transition. Through extensive finite-size scaling analysis, we show that this transition can truly be characterized as a second-order transition for which we have determined the critical exponents and established their relationship beyond mean-field methods. Practically, our protocol neither demands a sophisticated measurement scheme nor any specific initialization. A potential experimental verification, even if for a small effective atom number, can be in principle pursued by adapting the experimental setup put forward in~\cite{Ferioli2023}: in this work the evolution ruled by the BTC master equation~\eqref{eq:master-equation} and the corresponding phase-transition are observed for a cloud of laser-cooled Rubidium atoms in free space optically excited along its main axis. The total magnetization $\langle \hat{S}_z\rangle$, is then measured by collecting the emitted light in an avalanche photodiode. The measurement of the optimal spin-operator $\hat{S}_{\hat{\bf n}}$ described in our paper would thus need just an extra rotation of the atomic spin.

\hfill

\noindent \textbf{\large{Methods}}

\noindent \textbf{Ansatzes consistency} As discussed above, the quantum Fisher information $\mathcal{F}_Q$ should satisfy the following ansatz:
\begin{equation}
\mathcal{F}_Q(\omega) = \frac{a}{N^{-b}+c\left(\frac{\omega - \omega_\mathrm{max}(N)}{\kappa}\right)^\eta},
\label{eq_S:ansazt1}
\end{equation}
for a set of constants $a, b, c$ and $\eta$. Eq.~\eqref{eq_S:ansazt1} satisfies the scaling of the quantum Fisher information for large but finite system size $N\gg 1$ at $\omega = \omega_\mathrm{max}$ and the scaling in the thermodynamic limit $N\rightarrow \infty$. Factorizing Eq.~\eqref{eq_S:ansazt1} by $N^{b}$, one gets:
\begin{eqnarray}
\mathcal{F}_Q(\omega) &=& \frac{aN^{b}}{1+cN^{b}\left(\frac{\omega - \omega_\mathrm{max}(N)}{\kappa}\right)^\eta},\\
&=& \frac{aN^{b}}{1+c\left[N^{\frac{b}{\eta}}\left(\frac{\omega - \omega_\mathrm{max}(N)}{\kappa}\right)\right]^\eta},\\
&=& N^{b}f\left(N^{\frac{b}{\eta}}\frac{(\omega-\omega_\mathrm{max}(N))}{\kappa}\right).\label{eq:equality}
\end{eqnarray}
The above Eq.~\eqref{eq:equality} is of the same form as the one determined by the finite-size analysis
\begin{equation}
\mathcal{F}_Q(\omega)=N^{\frac{\eta}{\nu}}f\left(N^{\frac{1}{\nu}}\frac{(\omega-\omega_c)}{\kappa}\right).
\end{equation}
Hence, for consistency purposes, both ansatzes must be equal. By direct comparison, and assuming $\omega_\mathrm{max} = \omega_c$,  one obtains:
\begin{equation}
b = \frac{\eta}{\nu}
\end{equation}
which settles that the relationship between the three critical exponents $b, \eta$, and $\nu$ are not independent. Our numerical simulations perfectly match both independent analyses, showing that $\eta/\nu = 1.345 \pm 0.06$ from finite-size scaling is very close to the exponent $b=1.345$ obtained from studying the peak of the quantum Fisher information as a function of the system size $N$.

\hfill

\noindent \textbf{Time-constrained sensing bound} 

\noindent We now address the scenario where the time needed to perform the estimation protocol is somehow constrained, and thus the total time $T$ is also considered a resource. This is a standard framework in the context of frequency estimation, and one proves that in this case, the proper figure of merit to be optimized is the ratio between the QFI and the total evolution time, i.e. $\mathcal{F}_Q(\omega)/T$~\cite{Chin2012,Chaves2013,Brask2015,Smirne2016,Albarelli2018restoringheisenberg,Rossi2020PRL}.

As the dynamics encoding our parameter of interest $\omega$ is ruled by the Markovian master equation~(\ref{eq:master-equation}), a bound on the maximum QFI, corresponding to a quantum state evolved up to a time $T$, also considering possible adaptive control strategies, can be analytically obtained~\cite{Rafal2017}. 
In particular, one can write
\begin{align}
\mathcal{F}_Q(\omega) \leq 4  \max \lVert \hat{\alpha} \rVert T,
\end{align}
where $\lVert \hat{A} \rVert$ denotes the operator norm, and the operator $\hat{\alpha}$ reads
\begin{align}
\hat{\alpha} = |\gamma_1|^2 \,\hat{\mathbb{I}} + \gamma_1\gamma_2 \sqrt{\frac{\kappa}{S}} (\hat{S}_+ + \hat{S}_-) + |\gamma_2|^2 \frac{\kappa}{S} \hat{S}_+ \hat{S}_- \,.
\end{align}

In particular $\lVert \hat{\alpha} \rVert$ has to be maximized over the parameters $\{\gamma_j\}$ that satisfy the relation
\begin{align}
    \hat{S}_x + \gamma_1 \sqrt{\frac{\kappa}{S}} (\hat{S}_+ + \hat{S}_-) + \gamma_2 \frac{\kappa}{S} \hat{S}_+ \hat{S}_-  + \gamma_3 \hat{\mathbb{I}} = 0 \,. \label{eq:beta}
\end{align}
In order to fulfill Eq.~(\ref{eq:beta}) it is straightforward to observe that one has to fix $\gamma_2 = \gamma_3 = 0$ and 
$$
\gamma_1 = - \frac{1}{2} \sqrt{\frac{S}{\kappa}} \,.
$$
As a consequence one has that $\lVert \hat{\alpha} \rVert = (S / 4\kappa)$ needs no further optimization and, as $S=N/2$, the upper bound on the ration between the QFI $\mathcal{F}_Q(\omega)$ and the evolution time $T$ reads
\begin{align}
\frac{\mathcal{F}_Q(\omega)}{T} \leq \frac{N}{2\kappa} \,. \label{eq:DDbound}
\end{align}

We have then demonstrated, that whenever the time $T$ is considered a resource, one will eventually observe a linear scaling in the system size $N$, no matter the values of the parameter $\omega$ and $\kappa$ ruling the evolution. 

\hfill

\noindent \textbf{\large{Data availability}}

\noindent The data that support the ﬁndings of this study can be provided upon reasonable request.

\hfill

\noindent \textbf{\large{Code availability}}

\noindent The codes for analyzing the data of this study are available online at https://github.com/vm-physics/Boundary-Time-Crystals.

\hfill

\noindent \textbf{\large{Author contributions}}

\noindent The idea has been suggested by M.G.A.P. and shaped through discussions among all the authors. Numerical simulations have been performed by V.M. All the authors contributed in writing the paper.

\hfill

\noindent \textbf{\large{Acknowledgements}}

\noindent A.B. acknowledges support from the National Key R$\&$D Program of China (Grant No. 2018YFA0306703), the National Science Foundation of China (Grants No. 12050410253, No. 92065115, and No. 12274059), and the Ministry of Science and Technology of China (Grant No. QNJ2021167001L). V.M. thanks the National Natural Science Foundation of China (Grant No. 12050410251) and the Postdoctoral Science Foundation of China (Grant No. 2022T150098)).

\hfill

\noindent \textbf{\large{Competing interests}}

\noindent The authors declare no competing interests.

\hfill

\hfill

\bibliography{BTC_Bibliography}

\end{document}